\documentstyle[preprint,aps]{revtex}  
\begin{document}

\preprint{KNUTH-31}
\draft


\newcommand{\nc}{\newcommand}
\newcommand{\rnc}{\renewcommand}


\nc{\ignore}[1]{}

\nc{\be}{\begin{equation}}
\nc{\ee}{\end{equation}}
\nc{\bea}{\begin{eqnarray}}
\nc{\eea}{\end{eqnarray}}

\rnc{\a}{\alpha}
\rnc{\b}{\beta}
\rnc{\d}{\delta}
\nc{\e}{\eta}
\nc{\eb}{\bar{\eta}}
\nc{\f}{\phi}
\nc{\fb}{\bar{\phi}}
\nc{\vf}{\varphi}
\nc{\p}{\psi}
\rnc{\pb}{\bar{\psi}}
\rnc{\c}{\chi}
\nc{\cb}{\bar{\c}}
\rnc{\l}{\lambda}
\nc{\m}{\mu}
\nc{\n}{\nu}
\nc{\s}{\sigma}
\rnc{\o}{\omega}
\rnc{\t}{\theta}
\nc{\T}{D_0}
\nc{\bt}{}
\nc{\tb}{\bar{\theta}}
\nc{\ep}{\epsilon}
\nc{\im}{\imath}
\nc{\tc}{\tilde{C}}
\nc{\tB}{\tilde{B}}
\nc{\tJ}{\tilde{J}}
\nc{\te}{\tilde{\eta}}



\nc{\trac}[2]{{\textstyle\frac{#1}{#2}}}

\nc{\mat}[4]{\left(\begin{array}{cc}#1&#2\\#3&#4\end{array}\right)}

\def\tr{\mathop{\rm tr}\nolimits}
\def\Tr{\mathop{\rm Tr}\nolimits}
\nc{\r}{\rightarrow}
\nc{\ot}{\otimes}
\rnc{\ss}{\subset}
\nc{\ra}{\,\rangle}
\nc{\la}{\langle\,}
\nc{\ad}{A^\dagger}
\nc{\bd}{B^\dagger}
\nc{\ud}{U^\dagger}
\nc{\rf}[1]{(\ref{#1})}
\rnc{\ll}{\label}
\nc{\del}{\partial}
\rnc{\=}{ \; = \; }
\rnc{\-}{ \: - \: }
\rnc{\+}{ \: + \: }
\nc{\bs}{ B^s_+ \, B^s_- }
\nc{\bw}{ B^w_+ \, B^w_- }
\rnc{\sb}{ B^s_- \, B^s_+ }
\nc{\wb}{ B^w_- \, B^w_+ }
\nc{\qnb}{\{ \, n_b \, \} }
\nc{\eq}[1]{(\ref{#1})}
\nc{\com}[3]{[ \: #1 \: ,\: #2 \: ] &=& #3}
\nc{\tcom}[3]{[ \: #1 \: ,\: #2 \: ] \= #3}

\nc{\CV}{Calogero-Vasiliev}
\nc{\os}{oscillator}
\nc{\ie}{{\em i.e.}}
\nc{\HW}{Heisenberg-Weyl}
\nc{\HWA}{Heisenberg-Weyl algebra}
\nc{\eod}{\end{document}}

\title{ Solvable Potentials from Supersymmetric Quantum Mechanics}
\author{ Dong Sup Soh and Kyung-Hyun Cho} 
\address{
Department of Physics\\
Chonbuk National University\\
Chonju, \hspace{.1cm} {\bf 560-756},\hspace{.2cm} {\bf Korea}}
\author{Sang Pyo Kim} 
\address{
Department of Physics\\
Kunsan National University \\
Kunsan, \hspace{.1cm} {\bf 573-701},\hspace{.2cm} {\bf Korea}}
\date {July 27, 1995}
\maketitle
\begin{abstract}
A recurrence relation of Riccati-type differential equations
known in supersymmetric quantum mechanics is investigated
to find exactly solvable potentials.
Taking some simple {\it ans\"atze},
we find new classes of solvable potentials as well as
reproducing the known shape-invariant ones.
\end{abstract}
\vskip 1in
\centerline{( Submitted to Journal of Korean Physical Society )}
\newpage



It is generally difficult to solve exactly the
eigenvalue problem of a (time-independent) hamiltonian in quantum mechanics.
Among the various methods developed for this purpose,
a remarkably simple but powerful one is a ladder operator technique,
whose typical example is a simple harmonic oscillator.
If a hamiltonian has a discrete eigenvalue spectrum bounded from below,
the energy eigenstates should be labeled by integers
and formal rasing and lowering operators can be written in this basis.
However, it does not provide a way to find the ladder operators explicitly
for a given hamiltonian.
A practical method of getting ladder operators has been studied
by several authors(\cite{gend}-\cite{chuan})
using ideas of supersymmetric quantum mechanics(\cite{wit}, \cite{cooper})
and a concept of shape-invariant potentials \cite{gend}.
This approach has (re)produced many exactly solvable potentials.
In this paper, we investigate the recurrence relation of Riccati-type
differential equations arising in the supersymmetric quantum mechanics
directly taking some {\it ans\"atze}. In addition to reproducing known
shape-invariant potentials concisely, new classes of solvable potentials
are found.

Let us begin by summarizing some results of supersymmetric quantum mechanics
which are used for the present work(for a review, see \cite{cooper}
for example). Consider a family of one dimensional hamiltonians
\be
H_k \= \frac{p^2}{2} + V_k( x ) \;,\;\;\;\;\;\;\;\;\;
k = 0, 1, 2, \cdot\cdot\cdot \:,
\ll{hkx}
\ee
where $p \= - i \frac{\partial}{\partial x} \:,$
which are connected by
\be
H_{k+1} \= B_k \bd_k - E_k( 1 ) \= \bd_{k+1} B_{k+1}
\;,\;\;\;\;\;\;\;\;\;\;\;\;
H_0 \= \bd_0 B_0 \;.
\ll{hk}
\ee
Here $B_k$ and $\bd_k$ are defined using superpotentials $\a_k( x )$
\be
B_k \= \frac{1}{\sqrt2} \{ \frac{\partial}{\partial x} + \a_k( x ) \}
\;,\;\;\;\;\;\;\;\;\;
\bd_k \= \frac{1}{\sqrt2} \{ - \frac{\partial}{\partial x} + \a_k( x ) \} \:,
\ll{bkx}
\ee
where $\a_k( x )$ is a real function such that
the ground state wavefunction which is proportional to
$exp( - \int \a_k( x ) dx )\:$ should be normalizable.
The $n$th eigenvalue of $H_k$ is denoted by $E_k(n) \:$ $n = 0, 1, 2,
\cdot\cdot\cdot \:, $ and a relation between eigenvalues is given by
\be
E_{k+1}( n ) \= E_k( n + 1 ) - E_k( 1 ) \= E_0( n + k + 1 ) - E_0( k+ 1 ) \;.
\ll{energyk}
\ee
Note that all the ground state energies are defined to be zero.
The $E_k( 1 )$ is assumed to be positive and is also denoted by $s_k$,
standing for a level spacing between the $k$th and $(k + 1)$th
eigenstate of the starting hamiltonian $H_0 $.
It is worth noting that
$E_0( n + 1 ) = \sum_{k = 0}^{ n } s_k \;,
\:  n = 0, 1, 2, \cdot\cdot\cdot  \:$ and $E_0( 0 ) = 0 \:.$
Denoting the $n$th eigenfunction of $H_k$ as $\p_{k(n)}( x )$,
important relations between the eigenfunctions follow from \eq{hk} as
\bea
\p_{k+1(n)}( x ) &=& \{ E_k( n + 1 ) \}^{-1/2} \: B_k \p_{k(n+1)}( x )
\;, \nonumber \\
\bd_k \: \p_{k+1(n)}( x ) &=& \{ E_k( n + 1 ) \}^{1/2} \p_{k(n+1)}( x ) \;.
\ll{bla}
\eea
Note that $B_k \: \p_{k(0)}( x ) = 0 \;.$
A repeated use of this relation gives
\be
\bd_0 \bd_1 \cdot\cdot\cdot \bd_k \: \p_{k+1(n)}( x )
\= \: \{ E_0( n + k + 1 ) E_1( n + k ) \cdot\cdot\cdot E_k( n + 1 ) \}^{1/2}
\: \p_{0(n+k+1)}( x ) \:.
\ll{ladd}
\ee
Putting $n = 0$, this equation can be used to get the excited-state
wavefunctions of $H_0$ from the knowledge of
the ground states of systems related as superpartners.

Now, putting $B_k$ and $\bd_k$ in \eq{bkx} into \eq{hk},
a recurrence relation of Riccati-type differential equations is obtained as
\be
2 V_{k+1}( x ) \= \a^2_{k+1}( x ) - \a^\prime_{k+1}( x )
\= \a^2_{k}( x ) + \a^\prime_{k}( x ) - 2 s_k \:,
\ll{ric}
\ee
where the prime means a derivative with respect to $x \:.$
To search for the possible family of $\a_k( x )$ and $s_k$
satisfying the difference differential equation \eq{ric},
it is convenient to put $\a_k( x )$ and $\a_{k+1}( x )$ as
\be
\a_k( x ) \= \frac{1}{2 f^\prime_k( x )} \{ 2 s_k f_k( x ) +
f_k^{\prime\prime}( x ) \} \;,\;\;\;\;\;\;\;
\a_{k+1}( x ) \= \frac{1}{2 f^\prime_k( x )} \{ 2 s_k f_k( x ) -
f_k^{\prime\prime}( x ) \} \:,
\ll{alpak}
\ee
which solve \eq{ric} identically.
It holds for non-zero $s_k$ only,
and in general, there can be added a free parameter
$\sigma_k$, say, inside the curly brackets in \eq{alpak}.
Note that $f_k( x )$ is proportional to
the first excited-state wavefunction divided by
that of the ground state of the $k$th hamiltonian.
The consistency in the two equations of \eq{alpak} requires
\be
\{ f^\prime_k( x ) f^\prime_{k+1}( x ) \}^\prime \=
2 \{ s_k f_k( x ) f^\prime_{k+1}( x )
\- s_{k+1} f^\prime_k( x ) f_{k+1}( x ) \} \:.
\ll{ricf}
\ee

We further rewrite equations \eq{alpak}, \eq{ricf}
by putting $f_k( x )$ as following :
\be
f_k( x ) \= \tilde{g}_k( x ) \: exp \,( \int \tilde{h}_k( x ) dx \,) \;.
\ll{sfk0}
\ee
Note that there is an ambiguity in the choice of $\tilde{h}_k( x )$,
or equivalently in $\tilde{g}_k( x )$ :
Adopting $g_k( x ) = \tilde{g}_k( x ) \, exp \,( \int \b_k( x ) dx \,) \;,$
and $h_k( x ) = \tilde{h}_k( x ) - \b_k( x ) \:,$
instead of tilded functions,
$f_k( x )$ remains the same for arbitrary $\b_k( x )$.
We remove this ambiguity by choosing $\b_k( x )$ such that
$h_0( x ) = 0 \;,$ and
\be
f_k( x ) \= g_k( x ) \: exp \,( \int h_k( x ) dx \,) \;,\;\;\;\;\;\;\;\;\;
f_k^{\prime} ( x ) \= f_0^{\prime} ( x ) \: exp \,( \int h_k( x ) dx \,) \;.
\ll{sfk}
\ee
The consistency in this procedure requires
for all $k = 0, 1, 2, \cdot\cdot\cdot \;,$
\be
f_0^{\prime} ( x ) \= g_k^{\prime} ( x ) \+ g_k( x ) \: h_k( x ) \;,
\ll{sf}
\ee
then the difference differential equation \eq{ric} is satisfied.
Now, equation \eq{ricf} becomes
\be
f_0^{\prime\prime} ( x ) \+ \frac{1}{2} \: f_0^{\prime} ( x )
\,( \, h_k( x ) + h_{k + 1}( x ) \,) \=
s_k g_k( x ) \- s_{k+1} g_{k+1}( x ) \:.
\ll{sricf}
\ee
Solving this equation for $g_k( x )$ and inserting it into \eq{sf} yields
\be
k f_0^{\prime\prime\prime} ( x ) \+  p_k ( x ) f_0^{\prime\prime} ( x ) \+
( \, q_k ( x ) + s_k - s_0 \, ) \, f_0^{\prime} ( x ) \-
s_0 h_k ( x ) f_0 ( x ) \= 0 \:,
\ll{dde}
\ee
where
\be
p_k ( x ) \= k h_k ( x ) \+ \t_k ( x ) \;,\;\;\;\;\;\;\;\;\;
q_k ( x ) \= \t_k^{\prime} ( x ) \+ h_k ( x ) \t_k ( x ) \;,
\ll{pqx}
\ee
and
\be
\t_k ( x ) \= \frac{1}{2} \: \sum_{j=0}^{k-1} \;
( \; h_j ( x ) + h_{j+1} ( x ) \; )
\;,\;\;\;\;\;\;\;\;\; \t_0 ( x ) \= 0 \:.
\ll{thetak}
\ee
Since \eq{dde} is identically satisfied for $k = 0$, it is
equivalent to the difference of $(k+1)$th and $k$th expressions
in \eq{dde}. Thus, with the difference denoted by
$\Delta y_k \equiv y_{k+1} - y_k$, we finally obtain
\be
f_0^{\prime\prime\prime} ( x ) \+ \Delta p_k ( x ) f_0^{\prime\prime} ( x )
\+ (\, \Delta q_k ( x ) + \Delta s_k \,) \, f_0^{\prime}( x )
\- s_0 \Delta h_k ( x ) f_0 ( x ) \= 0 \:.
\ll{sdde}
\ee
It is equivalent to the difference differential equation \eq{ric},
and we will try to find $f_0(x)$ and $s_k$ consistently
by making simple {\it ans\"atze} on $h_k(x)$ below.
The superpotential in \eq{alpak} is now given as
\be
\a_k( x ) \=  \frac{1}{f_0^{\prime} ( x )} \,
\{ \, s_0 f_0(x) \- ( \, k - \frac{1}{2} \, ) \, f_0^{\prime\prime}( x ) \, \}
\+ \frac{1}{2} \, h_k(x) \- \theta_k(x)  \;.
\ll{salpak}
\ee
Note that an overall constant factor of $f_0(x)$ does not affect
the superpotential.

Let us solve \eq{sdde}. The simplest {\it ansatz} on $h_k(x)$ may be putting
$h_k(x) = 0$ for all $k = 0, 1, 2, \cdot\cdot\cdot \:.$
In this case, \eq{sdde} becomes
\be
f_0^{\prime\prime\prime} ( x ) \+ \Delta s_k \, f_0^{\prime}( x )
\= 0 \:.
\ll{sd1}
\ee
This is solved consistently by putting
\be
\Delta s_k \= s \:,
\ll{sd2}
\ee
where $s$ is a constant parameter.
It is easy to find $f_0^{\prime}(x)$ in \eq{sd1},
and the superpotential is given in \eq{salpak} with $ h_k(x) = 0 \,,
\theta_k(x) =0$ in this {\it ansatz}, that is,
\be
\a_k( x ) \=  \frac{1}{f_0^{\prime} ( x )} \,
\{ \, s_0 f_0(x) \- ( \, k - \frac{1}{2} \, )
\, f_0^{\prime\prime}( x ) \, \} \;.
\ll{sp}
\ee
The eigenvalues are given by the common formula
\be
E_k( n ) = s_0 n + \frac{1}{2}\;s n ( n - 1 + 2k ) \;,
\ll{ek1}
\ee
referring to \eq{sd2} and \eq{energyk}. In the following,
we list the resulting superpotentials, while the potentials
can be written down explicitly by using $\: 2V_k(x) \=
\a^2_{k}( x ) - \a^\prime_{k}( x ) \;$ in \eq{ric}.
With constant parameters $b, d \,,$ and putting
$$ \e \= \sqrt {|s|} \;,\;\;\;\;\;\;\;\;\;\;
a_k \= \frac{2( \, s_0 + ks \,) - s}{2 \e} \;, $$
\begin{itemize}
\item[(1)] $s>0 \,$ ; $f_0^{\prime}(x) \= sin \, \e(x+d)$
\be
\a_k(x) \=
\left\{
\begin{array}{ll}
-a_k \, cot \, \e (x+d) \+ \frac{b}{sin \, \e (x+d)}
& \mbox{Eckart potential} \\
\frac{a_k + b}{2} \, tan \,  \frac{\e}{2} \, (x+d)
\- \frac{a_k - b}{2} \, cot \,  \frac{\e}{2} \, (x+d)
& \mbox{P\"oschl-Teller potential I}
\end{array}
\right.
\ll{pt1}
\ee
\item[(2)] $s<0 \,$ ; $f_0^{\prime}(x) \= c_1 \, sinh \, \e(x+d)
\+ c_2 \, cosh \, \e(x+d) \,$ with constant parameters $c_1, c_2 \;.$
Thus, the superpotential in \eq{sp} is
\be
\a_k(x) \= \frac{a_k \, ( \, c_1 + c_2 \, tanh \, \e (x+d) \,)}
{ c_2 + c_1 \, tanh \, \e (x+d)} \+ \frac{b}
{c_1 \, sinh \, \e(x+d) \+ c_2 \, cosh \, \e(x+d)} \;.
\ee
Some well-known potentials in the literature come out :
\item[a.] $f_0^{\prime}(x) \= sinh \, \e(x+d) \, \; (\, c_1=1 \,, c_2=0 \,)$
\be
\a_k(x) \=
\left\{
\begin{array}{ll}
a_k \, coth \, \e (x+d) \+ \frac{b}{sinh \, \e (x+d)}
& \mbox{Rosen-Morse potential} \\
\frac{a_k -b}{2} \, tanh \,  \frac{\e}{2} \, (x+d)
\+ \frac{a_k +b}{2} \, coth \,  \frac{\e}{2} \, (x+d)
& \mbox{P\"oschl-Teller potential II}
\end{array}
\right.
\ll{pt2}
\ee
\item[b.] Morse potential ; $f_0^{\prime}(x) \= cosh
\, \e (x+d) \, \; (\, c_1 = 0 \,, c_2 = 1 \,)$
\be
\a_k(x) \= a_k \, tanh \, \e (x+d) \+ \frac{b}{cosh \, \e (x+d)} \;.
\ee
\item[c.] Morse potential ; $ f_0^{\prime}(x) \= exp \, \e (x+d) \,,
\, ( \, c_1 = c_2 = \frac{1}{2} \, ) $
[ $ c_1 = -c_2 = -\frac{1}{2} \,$ also gives a similar potential.]
\be
\a_k(x) \= a_k \+ b \, exp \, ( \, -\e (x+d) \,) \;.
\ee
\item[(3)] $s=0 \,$ ; two cases arise as following.
\item[a.] Harmonic oscillator potential ; $f_0^{\prime}(x) \= 1$
\be
\a_k(x) \= s_0 \, ( \, x+d \,) \;.
\ee
\item[b.] Harmonic plus inverse-square potential ; $f_0^{\prime}(x) \=
x+d$
\be
\a_k(x) \= \frac{s_0}{2} \, (x+d) \- \frac{\l_k}{x+d} \;,\;\;\;\;\;\;\;\;
\l_k \= k \- \frac{1}{2} \+ b \;.
\ll{cal}
\ee
\end{itemize}
It should be noted that $k$ has a maximum allowed value when $ s<0$
since $s_k = s_0 + ks$ is assumed to be positive.
It is remarkable that a large part of the known exactly solvable potentials
given in refs. \cite{dabro}, \cite{levai1} are reproduced here
within a single {\it ansatz}. Further details of the potentials
and the wavefunctions can be found in refs. \cite{cooper}, \cite{dabro},
\cite{levai1}, \cite{cooper2} where a notion of shape-invariance of
potentials is utilized in the framework of supersymmetric quantum
mechanics.

Let us solve \eq{sdde} with an {\it ansatz}
which is slightly more general than assuming $h_k( x ) = 0 \;$
for all $k \;,$
\be
h_k ( x ) \=
\left\{
\begin{array}{ll}
0   & \mbox{for  k = even } \\
h( x )  & \mbox{ for  k = odd } \:.
\end{array}
\right.
\ll{try3}
\ee
Using for $k =$ even, odd integers respectively,
$$
\Delta h_k ( x ) = \pm h(x) \;,
\;  \Delta p_k ( x ) = h(x) \, \{ 1 \pm ( k + \frac{1}{2} ) \, \} \;,
\;  \Delta q_k ( x ) = \frac{h^{\prime}(x)}{2} +
   \frac{h^{2}}{4} \, \{ 1 \pm ( 2k + 1 ) \, \} \;,
$$
the equation \eq{sdde} reduces to
\bea
f_0^{\prime\prime\prime} ( x ) \+ \frac{3}{2} h(x) f_0^{\prime\prime} ( x )
\+ \, \{ \frac{h^{\prime}(x)}{2} + \frac{h^{2}(x)}{2} + \Delta s_0 \, \} \,
f_0^{\prime}( x ) \- s_0 h(x) f_0 ( x ) &=& 0 \;, \nonumber \\
f_0^{\prime\prime\prime} ( x ) \- \frac{1}{2} h(x) f_0^{\prime\prime} ( x )
\+ \, \{ \frac{h^{\prime}(x)}{2} - \frac{h^{2}(x)}{2} + \Delta s_1 \, \} \,
f_0^{\prime}( x ) \+ s_0 h(x) f_0 ( x ) &=& 0 \;,
\ll{s12}
\eea
for $k = 0, 1$ and
\be
f_0^{\prime\prime} ( x ) \+ \{ \, \frac{h(x)}{2} + \frac{w}{2h(x)} \, \} \,
f_0^{\prime}( x ) \= 0 \;,
\ll{s3}
\ee
which results from the consistency of \eq{sdde} for all $k = 2, 3, 4, 5,
\cdot\cdot\cdot \,, $
together with $\Delta s_k$ determined with a constant $w$ as
\be
\Delta s_{2k} \= \Delta s_{0} \+ k w \;,\;\;\;\;\;\;\;\;\;
\Delta s_{2k +1} \= \Delta s_{1} \- k w \;.
\ll{s4}
\ee
Combining equations \eq{s12} and \eq{s3}, one finally arrives at
\be
h^{\prime}( x ) \+ \frac{1}{2} h^{2}(x) \, ( \,\frac{1}{\s} + 1 \, )
\+ \l \s \= 0 \;,\;\;\;\;\;\;\;\;\;
2\s f_0^{\prime} ( x ) \= h(x) f_0 ( x ) \;,\;\;\;\;\;\;\;\;
w \= 2 \l \s \;,
\ll{s5}
\ee
where $\l, \s$ are
\be
\l \= \Delta s_0 + \Delta s_1 \;,\;\;\;\;\;\;\;\;\;
\s \= \frac{\Delta s_0 - \Delta s_1 - w}{4 s_0} \;.
\ll{s6}
\ee
We consider $\s \neq 0$, since otherwise
$h(x) = 0$ in \eq{s5}, which has been already covered.
The energy level spacings are now given as
\be
s_{2k} \= s_0 + k \l \;,\;\;\;\;\;\;\;\;\;
s_{2k + 1} \= ( \, 1 + 2 \s \,) \, \{ \, s_0 + ( k + \frac{1}{2} ) \, \l \, \}
\;. \ll{s7}
\ee
Here, $ \s > - 1/2 \: $ for positive level spacings.
 From this and \eq{energyk}, the energy levels are given as
\leftline{ $
E_k(n) \= ( 1+ \s ) \, n \, \{ \, s_0
\+ \frac{\l}{4} \, ( n - 1 + 2k ) \, \}  $ }
\be
\- (-1)^k \, \s \, \{ \, \frac{1- (-1)^n}{2} \, ( \, s_0 \+
\frac{\l}{4} \, ( 2k - 1 ) \, ) \- \frac{\l \, n}{4} \, (-1)^n \, \} \;.
\ll{ek2}
\ee
This formula reduces to \eq{ek1} as $\s \r 0 \;,$ by noting in \eq{s7},
$\Delta s_k = \l / 2$ in this case.
The superpotential in \eq{salpak} is given finally using \eq{s5}
\be
\a_k ( x ) \= \frac{2 \s}{h(x)} \, ( \, s_0 - \frac{\l}{4} +
\frac{\l k}{2} \, ) \, \- (-1)^{k} \, \frac{h(x)}{4} \;.
\ll{s9}
\ee

Now, we only need to find $h(x)$ satisfying \eq{s5}.
The resulting superpotentials are listed in the following with a constant
$d \,,$ and putting
$$ \e \= \sqrt { \frac{|\l| \, (\, 1+ \s \,)}{2}} \;,\;\;\;\;\;\;\;\;
b_k \= (\, 1+ \s \,) \,( \, s_0 \- \frac{\l}{4} \+ \frac{\l k}{2} \,)
\;,\;\;\;\;\;\;\;\;\;
\m_k \= \frac{\s \, (-1)^k}{2(\, 1+ \s \,)} \;. $$
\begin{itemize}
\item[(1)] $\l > 0 \,$ ; $h(x) \= \frac{2 \s \e}{1+ \s}
\, cot \, \e \, (x+d)$
\be
\a_k(x) \= b_k \, \frac{tan \, \e \, (x+d)}{\e} \- \m_k \, \e \,
cot \, \e \, (x+d) \;.
\ee
\item[(2)] $\l < 0 \,$ ; $h(x) \= \frac{2 \s \e}{1+ \s}
\, coth \, \e \, (x+d)$
\be
\a_k(x) \= b_k \, \frac{tanh \, \e \, (x+d)}{\e} \- \m_k \, \e \,
coth \, \e \, (x+d) \;.
\ee
\item[(3)] $\l = 0 \,$ ; $h(x) \= \frac{2 \s }{1+ \s}
\, \frac{1}{x+d}$
\be
\a_k(x) \= s_0 \, (1+ \s) \, (x+d) \- \frac{\m_k}{x+d} \;.
\ll{soh1}
\ee
\end{itemize}

These potentials are similar but different from the ones obtained
by our first {\it ansatz}, that is, P\"oschl-Teller potentials I, II
given in \eq{pt1}, \eq{pt2} respectively, and \eq{cal}.
This fact is clear in view of their different energy-level structures
and the wavefunctions. For example, the potentials corresponding to the
superpotentials in \eq{cal}, \eq{soh1} are given respectively
(putting $d=0$ for simplicity)
\be
V_k( x ) \= \frac{s_0^2}{8} \, x^2 \+
\frac{\l_k^2 - \l_k}{2 x^2}
\- \frac{s_0}{2} ( \l_k + \frac{1}{2} ) \:,
\ll{cal2}
\ee
\be
V_k(x) \= \frac{1}{2} s_0^2 (  1 + \s )^2 \, x^2 \+
\frac{\m_k^2 - \m_k}{2 x^2}
\- \frac{s_0}{2} \, ( \, 1 + \s + (-1)^k \s \,) \;.
\ll{soh2}
\ee
The energy level spacings are
$ s_k = s_0 $ for all $ k $, for the
potentials \eq{cal2}, while $ s_k = s_0 \,, s_0 \, ( \, 1 + 2\s \,) $
for $ k = $ even, odd integers respectively for \eq{soh2}.
The difference between the two systems can be seen more clearly
in their wavefunctions.
The unnormalized wavefunctions for $V_0(x)$ given in \eq{cal2} are
\be
\p_{0(k)}(x) \= x^{\l_0} \,
exp (\, - \frac{1}{4} s_0 x^2 \, )
\, L_k^{\l_0 - \frac{1}{2}} ( \, \frac{s_0 x^2}{2} \, ) \:,
\ll{cal3}
\ee
while for $V_0(x)$ given in \eq{soh2}
\bea
\p_{0(2k)}(x) &=& x^{\m_0} exp (\, - \frac{1}{2} s_0 ( 1 + \s ) x^2 \, )
\, L_k^{\m_0 - \frac{1}{2}} ( \, s_0 ( 1 + \s ) x^2 \, ) \;, \nonumber \\
\p_{0(2k+1)}(x) &=& x^{1-\m_0} \, exp (\, - \frac{1}{2} s_0 ( 1 + \s ) x^2 \, )
\, L_k^{ \frac{1}{2} - \m_0 } ( \, s_0 ( 1 + \s ) x^2 \, ) \:,
\ll{soh3}
\eea
where $L_n^a (z)$ is a generalized Laguerre polynomial.
Note that as $\s \r 0 $, the potentials in \eq{soh2}, the wavefunctions in
\eq{soh3} and the energy levels all reduce to those of the simple
harmonic oscillator. In contrast, if the parameter is chosen such that
the potentials in \eq{cal2} become quadratic, only `half' of the energy
levels and wavefunctions of a corresponding harmonic oscillator
are obtained as can be seen in \eq{cal3}.
We remark that the wavefunctions in \eq{soh3} are normalizable
as long as $\s > - 1/2 $, although their first derivatives are not
continuous at $x=0 \;.$
The point is that another class of potentials like in \eq{soh2}
is possible as well as the well-known ones like in \eq{cal2}
if we seek supersymmetry-related potentials possessing normalizable
wavefunctions.

How can the (radial) Coulomb potential consisting of $1/x$ and $1/x^2$
terms be reproduced in our approach? Since we need $h_k(x)$ as inputs
for solving \eq{sdde}, we assume $h_k(x)$ to be that of Coulomb potential
\be
h_k(x) \= g \, ( \, \frac{1}{k + a} \- \frac{1}{k + 1 + a } \, )
\- \frac{g}{ a(a+1)} \;,
\ll{cu1}
\ee
where $ g, a $ are constant parameters.
The derivation of this will be described later
together with other general idea of getting superpotentials.
Then \eq{sdde} becomes for $k=0 \,,$
\be
f_0^{\prime\prime\prime} ( x ) \+ \frac{3 h_1}{2} \,
f_0^{\prime\prime} ( x ) \+ \{ \, ( \, \l + \frac{g( 2a + 1)}{a(a+1)} \, )
\, h_1 \+ \frac{g^2}{a^2 (a+1)^2} \+ b \, \} \, f_0^{\prime}( x ) \-
s_0 h_1 \, f_0 ( x ) \= 0 \;,
\ll{cu2}
\ee
where $ h_1 \= - 2g/ \{ \, a(a+1)(a+2) \, \} \;,$ and
\be
(2a+1) \, f_0^{\prime\prime} ( x ) \-
\{ \, 2 \l \+ \frac{2g(2a+1)}{a(a+1)} \, \} \, f_0^{\prime}( x )
\+ 2 s_0 \, f_0 ( x ) \= 0 \;,
\ll{cu3}
\ee
which results from the consistency of \eq{sdde} for $k = 1, 2, 3,
\cdot\cdot\cdot \;.$ Here, $s_k$ is determined as
\be
s_k \=  \frac{g^2}{2} \, ( \, \frac{1}{(k+a)^2} \-
\frac{1}{(k+ 1 +a)^2} \,) \+ \l h_k \+ bk \+ c \;,
\ll{cu4}
\ee
with another constants $\l, b, c \:.$
Combining \eq{cu2} and \eq{cu3},
interesting cases arise only when
\be
\l \= 0 \;,\;\;\;\;\;\;\;
c \=  ( \, a + \frac{1}{2}  \,) \, b \;,\;\;\;\;\;\;\;
a \neq -\frac{1}{2} \;,
\ll{only}
\ee
and the final equation is
\be
f_0^{\prime\prime} ( x ) \- 2 B \, f_0^{\prime}( x )
\+ ( \, B^2 + b \, ) \, f_0 ( x ) \= 0 \;,
\ll{cu5}
\ee
where
$$
B \= \frac{g}{a(a+1)} \;.
$$
Now, energy levels are obtained from above and \eq{energyk} as
\be
E_k(n) \= \frac{g^2}{2} \,\{ \frac{1}{( k + a )^2}
\- \frac{1}{( k + n +a )^2} \} \+
b \, \{ \frac{n^2}{2} \+ (k + a) \, n \, \} \:.
\ll{cu6}
\ee
The superpotentials are obtained using \eq{salpak}, \eq{cu5}
\be
\a_k(x) \= ( \, k \+ a \, ) \, F(x) \+ \frac{g}{ k+a } \;,\;\;\;\;\;\;\;\;\;
F(x) \= ( B^2 + b ) \,\frac{f_0(x)}{f_0^{\prime}(x)} \- B \;,
\ll{cu7}
\ee
where $f_0(x)$ is easily obtained in \eq{cu5}. We list the resulting
superpotentials in the following with a constant $d$ (to be chosen
suitably if necessary), and putting
$$ \e \= \sqrt{ |b| } \;.$$
\begin{itemize}
\item[(1)] $ b > 0  $ ;
$ f_0(x) \= sin \, \e (x+d) \, exp \,( \, Bx \,) $
\be
\a_k(x) \= - (k +a) \, \e \, cot \, \e x \+ \frac{g}{k+a} \;.
\ee
\item[(2)] $ b < 0  $ ;
$ f_0(x) \= ( \, c_1 \, sinh \, \e x +
c_2 \, cosh \, \e x  \,) \, exp \,( \, Bx \,) $ ;
with appropriate choices of $c_1, c_2 \,,$
there result two well-known potentials
\item[a.] Eckart potential
\be
\a_k(x) \= - (k +a) \, \e \, coth \, \e x \+ \frac{g}{k+a} \;.
\ee
\item[b.] Rosen-Morse potential
\be
\a_k(x) \= - (k +a) \, \e \, tanh \, \e x \+ \frac{g}{k+a} \;.
\ee
\item[(3)] $ b = 0  $ ; Coulomb potential
; $ f_0(x) \= ( x+d) \, exp \, ( \, Bx \,) $
\be
\a_k(x) \= -\frac{k+a}{x} \+ \frac{g}{k+a} \;.
\ee
\end{itemize}
It is satisfying to see that a single {\it ansatz} reflecting the common
structure of superpotentials as in \eq{cu7} produces a class of potentials
in addition to the starting one, the Coulomb potential in this case.
The above potentials and the results of our first {\it ansatz}
given in \eq{pt1}-\eq{cal}, constitute the whole shape-invariant
potentials given in refs. \cite{dabro}, \cite{levai1}. (There are some
discrepancies compared with the tables presented in these references.)

Finally, we want to mention another way of getting superpotentials
as following. Since it is generally expected that
the ground state wavefunctions are products of two factors,
or equivalently $\a_k( x )$ is a sum of two parts,
one may try to find new solutions to the
difference differential equation \eq{ric} by adding an appropriately
chosen second part to an old one which is assumed to satisfy it.
As a simple {\it ansatz} to implement this idea, we take
\be
\a_k( x ) \= a_k \; F( x ) \+ b_k \; G( x ) \:.
\ll{try}
\ee
Here, simple forms are taken for the two parts with constants
$a_k , b_k $ and $k$-independent functions $ F( x ) , G( x ) \;,$
and $a_k F( x )$ is assumed to solve \eq{ric}.
Then inserting \eq{try} into \eq{ric} results in
\be
( a_k + a_{k+1} ) \; F^{\prime}( x )  \+ ( a_k^2 - a_{k+1}^2 ) \; F^{2}( x )
\= 2 s_k^{(a)} \;,
\ll{assf}
\ee
\be
( b_k + b_{k+1} ) \; G^{\prime}( x )\+
2 \; ( a_k b_k - a_{k+1} b_{k+1} ) \; F( x ) G( x ) \+
( b_k^2 - b_{k+1}^2 ) \; G^{2}( x ) \= 2 s_k^{(b)} \;,
\ll{assg}
\ee
together with a condition
\be
s_k \= s_k^{(a)} \+ s_k^{(b)} \;.
\ll{cond}
\ee
Recall that $s_k$ should be positive definite, while $s_k^{(a)},
s_k^{(b)}$ need not be so.
In fact, all the results obtained so far can also be obtained within
this simple {\it ansatz} as may be clear from the forms of the
obtained superpotentials, \eq{sp}, \eq{s9}, \eq{cu7}.
We illustrate the procedure by obtaining the Coulomb potential :
A solution to \eq{assf} corresponding to $s_k^{(a)} = 0$
is directly obtained to be proportional to $1/x$,
$a_k $ being determined correspondingly.
Next, note that \eq{assg} is consistently solved by taking
$a_k b_k$ to be independent of $k$
and $G(x)$ to be a constant, say $g \:.$
Then one may get
$$
\a_k (x) \= - \frac{k+a}{x} \+ \frac{g}{k+a} \;,\;\;\;\;\;\;\;\;\;
V_k(x) \= -\frac{g}{x} \+ \frac{ (k+a)(k+a-1)}{2 x^2} \+
\frac{g^2}{2(k+a)^2} \;.
$$
One can find $h_k(x)$ by taking the difference of \eq{salpak},
\be
\Delta \a_k(x) \= - \frac{f_0^{\prime\prime} ( x )}{f_0^{\prime}( x )}
\- h_k(x) \;,
\ee
which can be used to get the $f_0(x)$ and $h_k(x)$
for the known superpotentials by recalling $h_0(x) = 0 \,$.

In summary, we have investigated the recurrence relation of
Riccati-type differential equations in supersymmetric quantum mechanics.
Recasting the problem into \eq{sdde} and
taking some simple {\it ans\"atze} on $h_k(x)$,
we reproduced the known shape-invariant
potentials in refs. \cite{dabro}, \cite{levai1}, and obtained
another classes of potentials which look similar but are different
from the known ones, as illustrated in \eq{cal2}-\eq{soh3}.
It should be stressed that different types of potentials
can be obtained altogether within a common {\it ansatz}
on the superpotential as shown in the text.
By using more intricate {\it ans\"atze} or introducing some
new ideas into the formalism,
we expect that the method presented in this paper
will be helpful for finding out more solvable potentials,
as well as reproducing the known ones, e.g.,
Natanzon potentials(\cite{natan}, \cite{cooper2}, \cite{cooper}).

\acknowledgments

K. H. Cho is grateful to S. U. Park for helpful comments.
The present studies were supported in part
by the Basic Science Research Institute Program,
Ministry of Education, 1995, Project No BSRI-95-2434,
and by the Korea Science and Engineering Foundation
under Grant No 951-0207-056-2.

\newpage
\rnc{\Large}{\normalsize}

\eod
\begin{thebibliography}{99}
\addcontentsline{toc}{section}{References}
\frenchspacing
\small
\addtolength{\itemsep}{-4pt}

\bibitem{wit} E Witten, Nucl.Phys. {\bf B185}, 513 (1981)

\bibitem{cooper} F Cooper F, A Khare A and U P Sukhatme,
{\it preprint} LA-UR-94-569, hep-th/9405029,
{\it Supersymmetry and Quantum Mechanics} (1994)

\bibitem{gend} L E Gendenshtein, Zh. Eksp. Teor. Fiz. Pis. Red.
{\bf 38}, 299 (1983) ( Engl. Transl. JETP Lett. {\bf 38}, 356 (1983))

\bibitem{dabro} J W Dabrowska, A Khare A and U P Sukhatme,
J. Phys. A: Math. Gen. {\bf 21}, L195 (1988)

\bibitem{levai1} G Levai, J. Phys. A: Math. Gen. {\bf 22}, 689 (1989)

\bibitem{levai2} G Levai, J. Phys. A: Math. Gen. {\bf 25}, L521 (1992)


\bibitem{cooper2} F Cooper, J N Ginocchio and A Khare,
Phys. Rev. {\bf D36}, 2458 (1987)

\bibitem{chuan} C X Chuan, J. Phys. A: Math. Gen. {\bf 24}, L1165 (1991)

\bibitem{natan} G A Natanzon, Vestnik Leningrad Univ. {\bf 10}, 22 (1971)
; Teor. Mat. Fiz. {\bf 38}, 146 (1979)



\end{thebibliography}
